\documentclass[apj]{emulateapj}

\usepackage{graphicx}

\newif\ifpreprint\preprinttrue

\newcommand{\ntab}[2]{ \multicolumn{1}{#1}{#2} }
\newcommand{\beq}   { \begin{eqnarray} }
\newcommand{\eeq}[1]{\label{#1}\end{eqnarray}}
\newcommand{\eeqn}{\end{eqnarray}}
\newcommand{\Frac}[2]{\frac{\displaystyle\strut #1}{\displaystyle\strut #2} }
\newcommand{\ds}{\displaystyle}
\newcommand{\eff}{\mbox{eff}}

\bibpunct{(}{)}{;}{a}{}{,}

\makeindex
\citeindextrue

\shorttitle{VERA 22~GHz Fringe Search Survey}
\shortauthors{Petrov et al.}
\received{}
\revised{}
\accepted{}

\begin{document}

\title{VERA 22 GHz fringe search survey}

\author{L. Petrov\altaffilmark{\dag}}
\affil{Mizusawa Astrogeodynamics Observatory, NAOJ, Mizusawa 023-0861, Japan}
\email{Leonid.Petrov@lpetrov.net}
\author{Tomoya Horota}
\affil{National Astronomical Observatory of Japan, Mitaka, Japan}
\email{tomoya.hirota@nao.ac.jp}
\author{Mareki Honma}
\affil{National Astronomical Observatory of Japan, Mitaka, Japan}
\email{honmamr@cc.nao.ac.jp}
\author{Katsunori M. Shibata}  
\affil{Mizusawa Astrogeodynamics Observatory, NAOJ, Mitaka, Japan}
\email{k.m.shibata@nao.ac.jp}
\author{Takaaki Jike}
\affil{Mizusawa Astrogeodynamics Observatory, NAOJ, Mizusawa 023-0861, Japan}
\email{jike@miz.nao.ac.jp}
\author{Hideyuki Kobayashi}
\affil{Mizusawa Astrogeodynamics Observatory, NAOJ, Mizusawa 023-0861, Japan}
\email{hideyuki.kobayashi@nao.ac.jp}
\altaffiltext{\dag}{on leave from NVI, Inc. / NASA GSFC}

\begin{abstract}

This paper presents results of a survey search for bright compact
radio sources at 22~GHz with the VERA radio-interferometer. Each source from
a list of 2494 objects was observed in one scan for 2 minutes. The purpose
of this survey was to find compact extragalactic sources bright enough
at 22~GHz to be useful as phase calibrators. Observed
sources were either a) within 6 degrees of the Galactic plane, or 
b) within 11 degrees from the Galactic center; or c) within 2 degrees 
from known water masers. Among the observed sources, 549 were detected, 
including 180 extragalactic objects which were not previously observed with 
the very long baseline interferometry technique. Estimates of the correlated 
flux densities of the detected sources are presented.

\end{abstract}


\section{Introduction}
\label{s:introduction}

  VLBI Exploration of Radio Astrometry (hereafter VERA) is a Japanese VLBI 
array dedicated to phase-referencing astrometry to explore the Milky Way 
Galaxy \citep{honma2000a,kobayashi2004}. VERA observes Galactic H$_2$O 
and SiO maser sources and determines their parallax and proper motions 
at the 10~microarcsec accuracy level. Based on the high-precision astrometry 
of $\sim$1000 maser sources, VERA will study the three-dimensional structure 
and dynamics of the Galaxy's disk and bulge, revealing the true shape 
of the bulge and spiral arms, its precise rotation curve and the distribution 
of dark matter.

  The VERA array consists of four 20-m diameter radio telescopes spread over 
Japan, covering baselines ranging from 1000 km to 2300 km. Unlike other VLBI 
stations, VERA telescopes are equipped with a dual-beam system 
\citep{kawaguchi2000}, with which one can simultaneously observe 
two sources separated from each other by up to a hundred times the 
telescope's primary beam size. With such a system, VERA can observe
a Galactic maser source and an extra-galactic calibrator source at the 
same time and perform relative astrometry of the Galactic maser source 
with respect to the extra-galactic calibrator, which serves as a position 
reference. Simultaneous observations of two sources, rather than switching 
observations with single-beam telescopes, are expected to be more effective
in canceling out the tropospheric fluctuations, which are the main source 
of position error in differential VLBI astrometry at 22 GHz or higher 
frequency. In fact, preliminary results from VERA's dual-beam system 
observations have already shown its high capability of phase-referencing 
to reach down to 10~microarcsecond level accuracy \citep{honma2003}, and 
now monitoring of several bright maser sources is on-going to produce 
initial results 
from the VERA.

\begin{figure}[ht]
  {\includegraphics[width=0.48\textwidth,clip]{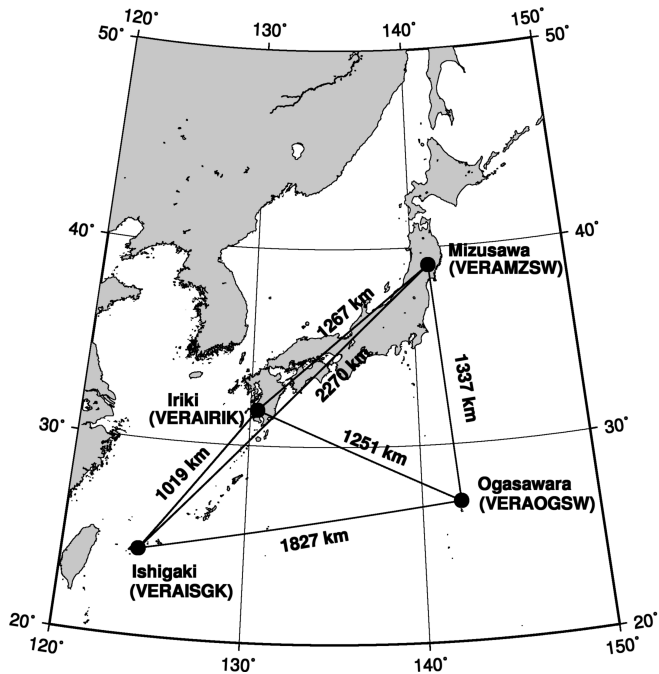} }
  \caption{VERA network.}
  \label{f:vera_network}
\end{figure}

  In order to observe a pair of two adjacent sources with VERA, the separation 
angle of the pair must be less than $2.\!{}^\circ2$. This is not only 
a mechanical limit of VERA's dual-beam system, but also a practical limit for 
effective phase referencing to achieve 10~microarcsec accuracy, i.e., 
the reference source needs to be close enough to the target maser to remove the 
tropospheric fluctuations sufficiently. Therefore, the existence of calibrators 
close to target sources is crucial to successful VERA observations. If we 
assume a uniform density of calibrators in the sky plane, 
$\sim$ 2700 calibrators are required in the whole sky to find a calibrator 
within $2.\!{}^\circ2 $ of any sky position (a circle with radius of 
$2.\!{}^\circ2 $ covers 0.0046 steradians in the sky; thus 
$4\pi / 0.0046 \approx 2700$ are necessary). Practically though, more sources 
are needed because the distributions over the sky is not uniform.

  To date, there have been many efforts to find good calibrators observable 
with VLBI. The pioneering work of \citet{first-cat} provided the first 
catalogue of source positions determined with VLBI, which contained 35 strong
sources. In 1998 the International Celestial Reference Frame (ICRF) catalogue 
of 667 sources produced by analyzing VLBI observations made in the framework 
of space geodesy programs was published by \citet{icrf98}. At present, the most
massive survey of compact calibrators is the VLBA Calibrator Survey, 
hereafter VCS, performed in 1994--2005 \citep{vcs1,vcs2,vcs3,vcs4}, and 
Kovalev~et~al. (submitted to \aj,\ 2006\footnote{Available at {\tt http://arxiv.org/astro-ph/0607524}}). After 
the fifth VCS campaign, the NASA catalogue 2005f\_astro\footnote{Available at 
{\tt http://vlbi.gsfc.nasa.gov/astro}} of all sources detected in the absolute 
astrometry mode at 2.2 GHz and/or 8.6 GHz frequency bands (S and X) contains 
3481 sources. Among them, 85\% are acceptable as calibrators for the VLBA. 
Nevertheless, in the sky area at $\delta \ge -40^\circ$, which is observable 
with the VERA, the probability of finding an X/S calibrator with correlated
flux density over 60~mJy within $2.\!{}^\circ2 $ of any position is $\sim$70\% 
\citep{vcs4}, indicating that more calibrators are needed in order 
to observe all maser sources with VERA. Moreover, maser sources tend to be 
confined to the Galactic plane, because they are basically located in the 
Galaxy's disk, but the density of known calibrators near the Galactic plane 
is even lower than that in other sky regions. This is due to the strategies 
used to find compact calibrators: the list of candidates was based on low 
resolution sky surveys that severely suffered from strong radio foreground 
in the Galactic plane. Besides, some past surveys, JVAS \citep{jvas}, VCS1, 
VCS3, VCS4, specifically avoided the Galactic plane. To overcome this problem, 
a calibrator survey in the Galactic plane has been done by \citet{honma2000b} 
finding more than 50 new calibrators at $|b| \le 5^\circ$, but still many more
calibrators in this region are needed.

  For these reasons, we have been performing the Galactic Plane Survey of 
compact radio sources, hereafter GaPS, based on the VERA and VLBA observations.
The objective of this campaign is to provide a list of calibrators for 
dual-beam VERA observations of H$_2$O masers at 22~GHz. Our strategy for 
GaPS is two-fold: first, we observe a wide list of candidate sources with 
the VERA, and then observe a narrow list of detected sources with the VLBA 
in order to determine their positions at the 1--10~nrad level of accuracy and 
to obtain high quality maps. In this paper, we present results from the 
first part of the GaPS project, the fringe search observations with the VERA, 
reporting the detection of 180 new compact sources. 
In section~\S\ref{s:observations} we describe our approach to the selection 
of candidates, scheduling and observations. In section~\S\ref{s:anal} we 
describe analysis of the observations. Results are discussed in 
section~\S\ref{s:results} and concluding remarks are given in 
section~\S\ref{s:summary}.

\section{Observations}
\label{s:observations}

\subsection{Source selection}

  We searched for compact extragalactic sources with correlated flux density
$> 100$~mJy at spacings $ \sim\! 10^{8} $ wavelengths at 22~GHz in the 
following zones: a) $ |b| < 6^\circ $  and $ \delta > -40^\circ$; 
b) within an $ 11^\circ $ radius of the Galactic center; and c) within 
$ 2.\!^\circ2 $ of objects listed in the 2nd update of Arcetri Catalog of 
H$_2$O maser sources \citep{h2o_masers} and the stellar masers catalogue 
of \citet{stellar_masers}. 

  First, we combed through the wide list of 3481 known compact sources observed 
under absolute astrometry programs at S/X bands. Among them, 510 fall in
our zones of interest. Also among the 3481 known compact sources, 252 objects 
have been observed under the K/Q band survey in 2002--2005 
\citep{fey_kq,jacobs_kq}, and can be considered as confirmed K-band 
calibrators. But only 16 of these fall in our zones of interest. The first list 
of targets was formed from the 510 known X-band calibrators, except 16 objects 
observed with the K/Q band survey. The purpose of including known X/S 
calibrators in the list of candidate sources was to check whether they are 
bright enough at K~band to be detected. It was expected that not all of these 
will be detected with the VERA, since although they have correlated flux 
densities above 60~mJy at X-band, their correlated flux densities at K-band 
may be below the VERA detection limit, and, therefore, not suitable as 
calibrators for VERA.

  Second, we searched the CATS database \citep{cats} containing almost all
radio catalogs\footnote{\url{http://cats.sao.ru/doc/CATS\_list.html}} to
find flux density measurements at radio frequencies 1.4~MHz and higher.
We selected the sources with at least two measurements of the flux density
and found their spectral index by fitting a straight line to the log(F)/log(f) 
dependence, where $F$ is the flux density and $f$ is the 
frequency. The flux density was extrapolated to 22~GHz. Those sources in 
the zone of our interest, which had extrapolated flux density $>100$~mJy 
at 22~GHz, and which were not observed in the absolute astrometry mode,
were put in the second list of targets, in total 1901 new sources. 

\subsection{Scheduling}

   The list of target sources was split into two group: the first group,
of 1386 objects, within $ 6^\circ$ of the Galactic plane or within $11^\circ$
from the Galactic center; and a second group, of 1032 sources, within 
$ 2.\!{}^\circ 2$ from known galactic maser sources. The first group had 
priority in scheduling. Among target sources, 94\% were observed in one scan
of 120~s, and 6\% were observed in two scans. Observing schedules were 
prepared using the following procedure. A source in the target list, which
was not observed in previous experiments and which had the highest elevation
at the station of VERZMZSW, was selected as the first objects in the schedule.
In order to select the next source, the pool of target sources was examined, 
and for each source, which was at least $10^\circ$ above the horizon and which 
was not previously observed, a score was computed: 
$ S = 100/T_{slew} + 30*( 20 - \delta ) + 0.2*E_{min} $ where $ T_{slew} $ 
is the slewing time in seconds, $\delta$ is the source declination, in degrees, 
and $ E_{min} $ is the minimum elevation of the source among the network 
stations, in degrees. The source from the first group with the maximum score was
put in the schedule. If no source from the first group was visible, the source 
with the highest score from the second group was put in the schedule. Then
the procedure was repeated. The slewing rate of VERA stations is 
$2^\circ\, s^{-1}$ and acceleration is $2^\circ\, s^{-2}$. The antenna control 
hardware requires adding a margin of 10 seconds to slewing time, and the 
slewing time should be rounded to the next 4~s. Measurement of system 
temperature takes an additional 30~seconds. System temperature was measured 
either every 5 scans or when the antennas moved more than $10^\circ$ from 
the previous measurement, whichever occurs first. In addition to targets, 
observations of amplitude calibrator sources were scheduled every hour. At 
the beginning and at the end of each session a very strong source, a fringe 
finder, was observed. The amplitude calibrator sources were taken from the 
list of 252 sources observed with the K/Q band survey. The average density 
of observations was 22 scans per hour.

\subsection{Observations}

  GaPS survey observations were scheduled mainly within gaps between regular
VERA observations (table~\ref{t:exp_table}). Some observations were performed 
only at three antennas while one of the stations was not available.

\begin{deluxetable}{l l r r r}
   \tablecaption{Observing sessions \label{t:exp_table}}
   \tablehead{
     \ntab{c}{(1)}  & 
     \ntab{c}{(2)}  & 
     \ntab{c}{(3)}  & 
     \ntab{c}{(4)}  & 
     \ntab{c}{(5)} 
    }
    \startdata
      r05284b   & 2005 Oct. 11 \enskip 08:20  &   7.0  &  3  &  1.0 \\
      r05285b   & 2005 Oct. 12 \enskip 08:20  &   7.0  &  4  &  1.0 \\
      r05332b,c & 2005 Nov. 28 \enskip 12:20  &  10.8  &  4  &  1.0 \\
      r05335b   & 2005 Dec. 01 \enskip 11:00  &  12.5  &  4  &  1.0 \\
      r05343b   & 2005 Dec. 09 \enskip 15:00  &   8.5  &  4  &  0.1 \\
      r05348a   & 2005 Dec. 14 \enskip 09:00  &  14.0  &  3  &  1.0 \\
      r05349a   & 2005 Dec. 15 \enskip 10:00  &  20.0  &  3  &  1.0 \\
      r05350a   & 2005 Dec. 16 \enskip 07:00  &   7.5  &  3  &  1.0 \\
      r05350b   & 2005 Dec. 16 \enskip 14:55  &   8.5  &  3  &  1.0 \\
      r05353d   & 2005 Dec. 19 \enskip 04:00  &   5.5  &  3  &  1.0 \\
      r05359d   & 2005 Dec. 25 \enskip 02:00  &   7.0  &  4  &  0.1 \\
      r06052b   & 2006 Feb. 21 \enskip 18:00  &  13.0  &  4  &  0.1 \\
      r06054a   & 2006 Feb. 23 \enskip 19:00  &  11.0  &  4  &  0.1 \\
      r06059b   & 2006 Feb. 28 \enskip 18:00  &  16.0  &  4  &  0.1 \\
   \enddata
   \tablecomments{ (1) session name; (2) start date; 
                   (3) duration in hours; (4) \# stations; 
                   (5) Accumulation period length in seconds. }
\end{deluxetable}

  The left circular polarization in band 21.97--22.47 GHz band was received,
sampled with 2~bit quantization, and filtered using the VERA digital 
filter \citep{iguchi2005} before being recorded onto magnetic tapes. The 
digital filter split the data within the 500 MHz band into 16 frequency 
channels of 16~MHz width each, equally spaced with spacing 32~MHz. The data 
were recorded with the SONY DIR2000 recorder at 1~Gbps rate.

\section{Data analysis}
\label{s:anal}
 
   Correlation was carried out on the Mitaka FX correlator 
\citep{chikada1991} with a spectral resolution of 250~kHz. Thus, the amplitude
and phase of the cross correlation function at 64 spectral channels within each
of the 16 frequency channels was determined. The correlator writes the output 
in the native CODA format. The output was re-formated to the FITS-IDI format 
(C.~Flatter, (1998) unpublished\footnote{Available on the Web at 
\hfill\hspace{1em}\linebreak
{\tt http://www.cv.nrao.edu/fits/documents/drafts/idi-format.ps}}). One part
of the observing sessions was correlated with accumulation periods of
0.998~s, and another part was correlated with accumulation periods 
of 0.1~s.

\begin{figure}[ht]
  {\includegraphics[angle=-90,width=0.48\textwidth]{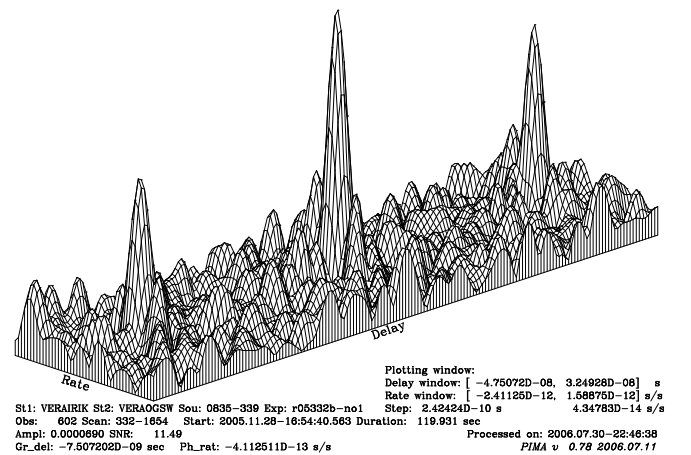} }
  \caption{Fringe amplitude versus delay and delay rate for a strong
           source.}
  \label{f:strong_det}
\end{figure}

\begin{figure}[ht]
  \vbox{\par\vspace{2ex}\par
          \includegraphics[angle=-90,width=0.48\textwidth]{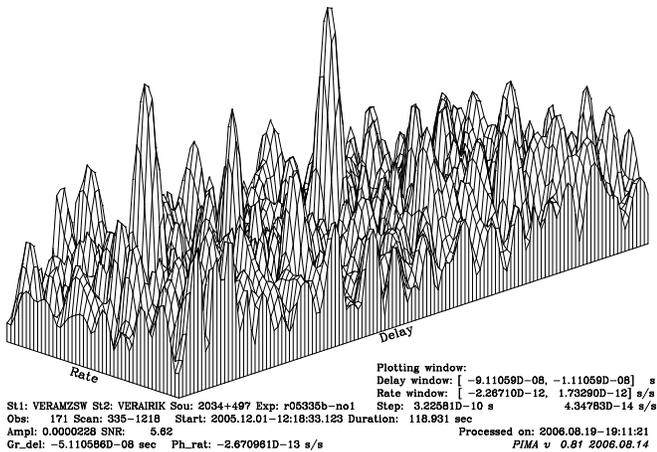} 
       }
  \caption{Fringe amplitude versus delay and delay rate for a marginally 
           detected source.}
  \label{f:weak_det}
\end{figure}

  The data were analyzed with program PIMA. The first step of analysis was
fringe fitting. The group delay and delay rate that maximize the fringe 
amplitude were sought for each baseline independently. The spectrum of the 
cross-correlation function was presented as a two-dimensional array, with
the first dimension running over frequency channels, and the second dimension
running over accumulation periods. The size of the matrix was 
$960 \!\times\! 120$ for scans with accumulation periods of 1~s and 
$1920 \!\times\! 1200$ for data with accumulation periods of 0.1~s. The 
grid was extended to $4096 \!\times\! 512$ and $8192 \!\times\! 4096$ for 
the data with 1~s and 0.1~s accumulation periods, respectively, with an
oversampling factor of four by padding extended elements with zeros. 
The amplitudes of the cross-correlation function was normalized by dividing 
it by the square root of the product of the autocorrelation of the recorded 
data at each station of the baseline. The two-dimensional fast Fourier 
transform was performed. The first dimension of the Fourier transform runs 
over group delays from 0 through 4~mks with steps of 0.49~ns, and the 
second dimension runs over delay rates from 0 through $ 4.6 \cdot 10^{-11} $ 
with steps of $ 0.9 \cdot 10^{-13} $ for the data with 1~s accumulation 
periods, and from 0 through $ 4.6 \cdot 10^{-10} $ with the step of 
$ 1.1 \cdot 10^{-13} $ for the data with the accumulation periods of 0.1~s. 
The coarse search for the maximum of the correlation amplitude was performed 
over results of the Fourier transform. In the worst case, when the maximum 
of the correlation amplitude falls just between the nodes of the grid, the 
amplitude is reduced by $\mbox{sinc}^2(\pi/(2v))$, where $v$ is the 
oversampling factor. When $v=4$, in the worst case the amplitude is reduced 
by $ \frac{16(2-\sqrt{2})}{\pi^2} = 0.95$. The oversampling is important for 
detecting weak sources. The fine search of the correlation maximum is 
performed by the consecutive direct Fourier transform at a grid 
$3 \!\times\! 3$ around the maximum with reducing spacings of the grid by 
half at each step of an iteration. Three iterations of the fine search were 
performed. Examples of fringe amplitude plots versus trial delay and delay rate
for a weak and a strong source are shown in 
figures~\ref{f:strong_det} and \ref{f:weak_det}.

\begin{figure}[ht]
  {\includegraphics[width=0.48\textwidth,clip]{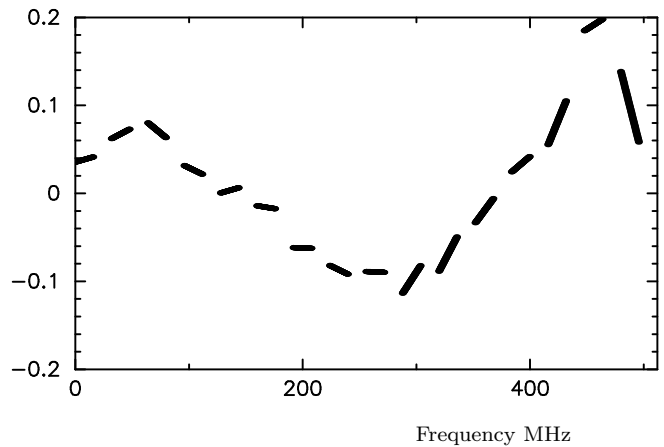} }
  \par\vspace{-0.1ex}\par\noindent\hspace{54mm}{Frequency MHz}
  \caption{The phase of the bandpass calibration at baseline
           \mbox{\sc veramzsw/veraogsw}.}
  \label{f:bps_phas}
\end{figure}

\begin{figure}[ht]
  {\includegraphics[width=0.48\textwidth,clip]{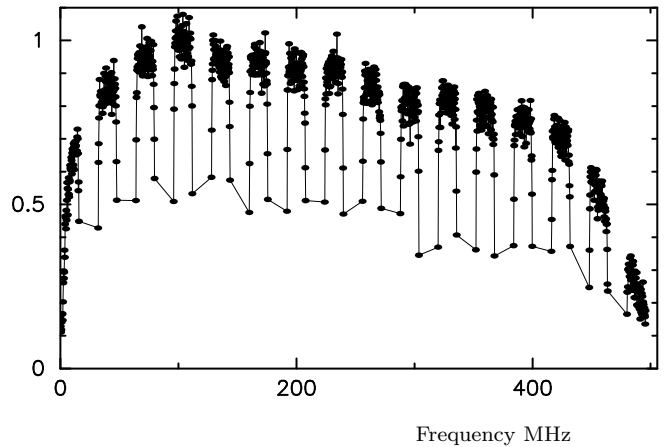} }
  \par\vspace{-0.1ex}\par\noindent\hspace{54mm}{Frequency MHz}
  \caption{The amplitude of the bandpass calibration at baseline
           \mbox{\sc veramzsw/veraogsw}.}
  \label{f:bps_ampl}
\end{figure}

  Observations of strong sources with signal to noise ratios exceeding 120 
were used for evaluation of a baseline dependent bandpass calibration. The 
phase of the bandpass calibration was determined as a residual phase with 
the opposite sign at each spectral channel after subtracting the multi-band 
fringe phase. A linear model was fitted into individual phases at spectral 
channels within each frequency channel. The amplitude of the band-pass was 
determined as the ratio of the fringe amplitude at an individual spectral 
channel to the maximum amplitude within the band. The estimate of the maximum 
amplitude across the band was found as the maximum value of the amplitude 
across 1024 spectral channels minus the doubled standard deviation of the 
amplitude at an individual spectral channel. It was found that the bandpass 
is stable between experiments, so the bandpasses over all experiments were 
stacked and averaged. An example of the bandpass calibration is shown in 
figures~\ref{f:bps_phas} and \ref{f:bps_ampl}. The complex bandpass calibration 
was applied to all observations, i.e. the $uv$ data were multiplied 
with the complex bandpass before the fringe search. This technique increased 
the fringe amplitude by 5--15\%.

\subsection{Calibration}

\begin{figure}[ht]
  \vbox{
        \includegraphics[angle=90,width=0.02\textwidth,clip]{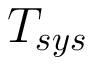}
        \includegraphics[width=0.45\textwidth,clip]{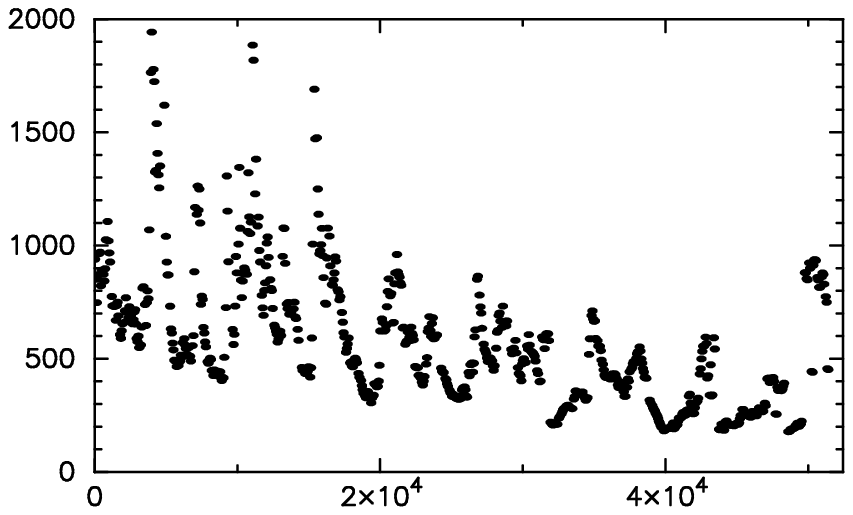}
        \par\vspace{-0.1ex}\par\noindent\hspace{70mm}\hspace{7mm}{Time s}
       }
  \caption{System temperature corrected for the atmosphere opacity versus 
           time from beginning of observing session at \mbox{\sc veraisgk} 
           on 2005.12.14.}
  \label{f:tsys_time}
\end{figure}

\begin{figure}[ht]
  \vbox{
        \includegraphics[angle=90,width=0.02\textwidth,clip]{petrov.fig12.eps}
        \includegraphics[width=0.45\textwidth,clip]{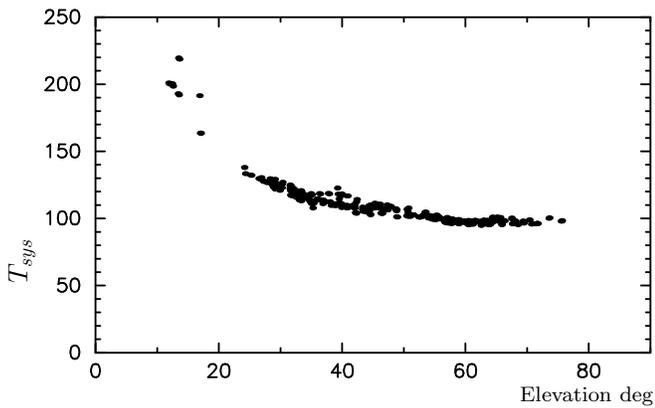}
        \par\vspace{-0.1ex}\par\noindent\hspace{58mm}\hspace{10mm}{Elevation deg}
       }
  \caption{System temperature corrected for the atmosphere opacity versus 
           elevation in degrees at \mbox{\sc veramzsw} on 2005.12.09.}
  \label{f:tsys_elev}
\end{figure}

  System temperatures including atmospheric attenuation were measured with 
the chopper-wheel method \citep{ulich1976}. A microwave absorber at 
ambient temperature was inserted just in front of the feed horn at the 
beginning of each scan, and the received total power was measured with 
a power meter. Using the measured total power for the blank sky and the 
absorber, we can determine the temperature scale automatically corrected 
for the atmospheric attenuation. The system temperatures during the best 
weather conditions were 100--200~K. However, at typical weather conditions, 
$T^\star_{sys}$ were 200-500~K depending on the elevation of observed sources, 
reaching 2000~K at low elevations and during adverse weather conditions. 
Figures \ref{f:tsys_time} and \ref{f:tsys_elev} illustrate recorded system 
temperatures at two stations at good and adverse weather conditions. 
The uncertainty in the temperature scale was estimated to be 10\%, mainly 
due to the assumption that the ambient temperature is the same as the 
air temperature 

  The aperture efficiency of each antenna was 45--52\% in the 22 GHz band, 
which was measured by observing the continuum emission from Jupiter once 
per year. The accuracy of the aperture efficiency measurement was typically 
2-5\%. The antenna gain does not show elevation dependency in the 
elevation range 10--$90^\circ$.

  Amplitude calibration was performed in two steps. First, the amplitude
conversion factor was computed for each observation based on measurements
of system temperature and antenna efficiency as 
$ A = \sqrt{ \frac{ T^\star_{1sys} \, T^\star_{2sys} }{g_1 \, g_2} } $ where 
$g_i$ is the antenna gain. This factor converts dimensionless fringe amplitude 
to correlated flux density. Then the multiplicative correction to the 
conversion factor was computed for each observation of the amplitude 
calibrator at each baseline as $ \delta A = \frac{F_{kq}}{F_{obs}} $  where
$ F_{kq} $ is the correlated flux density from maps from the KQ survey and 
$ F_{obs} $ is the observed correlated flux density. The correction 
$ \delta A $ was averaged over all amplitude calibrators within an individual
observing session. The scatter of the individual estimates of the correction
provided the measure of errors of amplitude calibration: 20--25\%. Since 
for the vast majority of sources, only 3 or 6 values of fringe amplitude and 
phase were determined, no attempt to make maps was made.
  
\subsection{Detection limit}

  Strong sources have a noticeable peak in plots of fringe amplitude versus
trial delay and delay rate, for instance figure~\ref{f:strong_det}. 
The presence of noise will force the fringe search process to find a peak in 
such a plot even in the absence of a signal from the source. For each 
observation we need to evaluate the probability of false detection. In the 
absence of signal the amplitude of the cross correlation function $z$ has 
Rayleigh distribution \citep{r:papolius}:
\beq
    p(z) = \Frac{z}{\sigma^2} e^{-(\frac{z^2}{2\sigma^2})}
\eeq{e:e1}
  where $\sigma$ is the standard deviation of the real and imaginary part of
the cross correlation function. The noise level of fringe amplitudes was 
evaluated for each observation. At the grid of cross correlation function 
spectrum 16384 points were randomly taken and the average amplitude was 
computed. In order to insure that no data points with signal were taken, 
an iterative procedure removed all the points which exceeded the average by
more than 4 times. The average amplitude of noise in the spectrum was 
considered as an estimate of the mathematical expectation of the noise, 
$ {\cal E}(z) = \sqrt{\pi/2}\, \sigma$. In the case when all points of the 
spectrum are {\it statistically independent}, the cumulative probability 
for the maximum of $n$ points is \citep{r:tms}
\beq
    P(z) = \left( 1 - e^{-(\frac{ z^2}{2\sigma^2})} \right)^n
\eeq{e:e2}
  Then, differentiating this expression, we find the probability distribution
density of the ratio of the maximal fringe amplitude to the averaged noise 
at points in the spectrum with no signal, $s = z_{max}/<\!\!z\!\!>$:
\beq
    p(s) = \Frac{2}{\pi} \Frac{n}{<\!\!z\!\!>} \, s \, e^{-\frac{s^2}{\pi}}
           \left( 1 - e^{-\frac{ s^2}{\pi}} \right)^{n-1}
\eeq{e:e3}

\begin{figure}[ht]
  {\includegraphics[width=0.48\textwidth,clip]{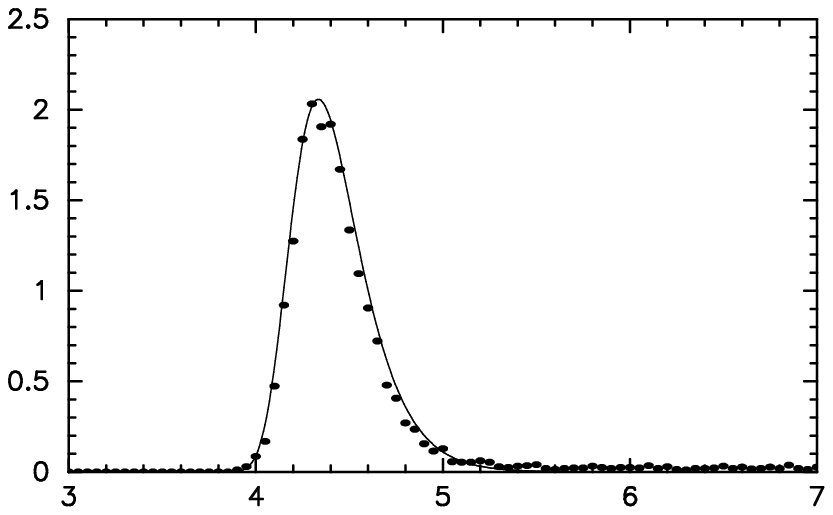} }
  \par\vspace{-2.0ex}\par\noindent\hspace{84mm}{\large s}
  \caption{The distribution density of the achieved signal to noise
           ratios (points) and the regression curve (continuous line)
     \mbox{$
             p(s) = \Frac{2}{\pi} \Frac{n_{\eff}}{<\!\!z_{\eff}\!\!>} \, s \, 
                    e^{-\frac{s^2}{\pi}}
                    \left( 1 - e^{-\frac{ s^2}{\pi}} \right)^{n_{\tiny\eff}-1}
           $}
     for scans with accumulation periods 1.0~s. 
     \mbox{$ <\!\!z_{\eff}\!\!> = 1.101, n_{\eff} = 183000$}. 
  }
  \label{f:prob_snr_long}
\end{figure}

\begin{figure}[ht]
  {\includegraphics[width=0.48\textwidth,clip]{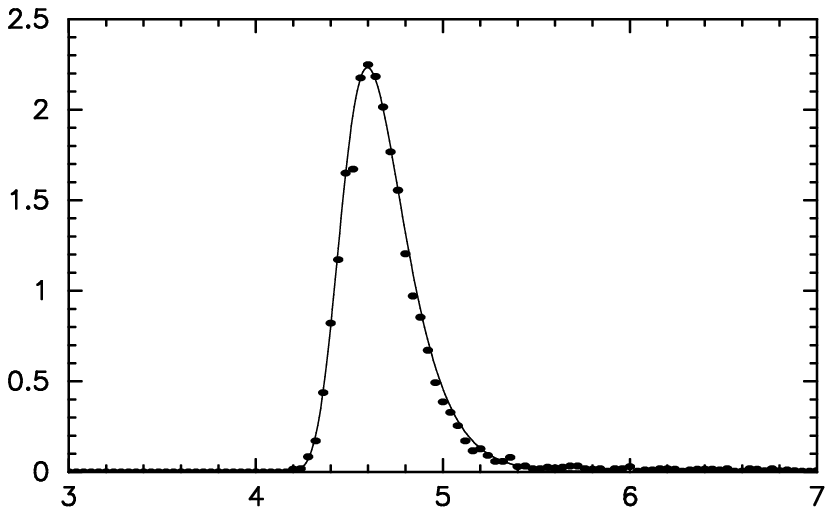} }
  \par\vspace{-2.0ex}\par\noindent\hspace{84mm}{\large s}
  \caption{The distribution density of the achieved signal to noise
           ratios (points) and the regression curve (continuous line)
     \mbox{$
            p(s) = \Frac{2}{\pi} \Frac{n_{\eff}}{<\!\!z_{\eff}\!\!>} \, s \, 
                   e^{-\frac{s^2}{\pi}}
                   \left( 1 - e^{-\frac{ s^2}{\pi}} \right)^{n_{\tiny\eff}-1}
           $}
     for scans with accumulation periods 0.1~s. 
     \mbox{$ <\!\!z_{\eff}\!\!> = 1.091, n_{\eff} = 989000$}.
  }
  \label{f:prob_snr_short}
\end{figure}

  It should be emphasized that expression \ref{e:e3} is valid if all
points of the spectrum are independent. This a~priori distribution density
is an approximation, which is valid only partially due to oversampling and 
deviation of the bandpass from a rectangular shape. Using the sample of 
achieved signal to noise ratios, we computed the a~posteriori distribution
density. We assumed that the a~posteriori distribution can be approximated
as a function like expression~\ref{e:e3} with effective parameters 
$<\!\!z\!\!>_{\eff}$ and $n_{\eff}$. We found parameters of the distribution 
which minimize residuals between the a~priori and a~posteriori distribution 
densities in the range of signal to noise ratios [3,\, 6.5] by searching 
through the space of trial values using the brute force algorithm. Plots 
of the a~posteriori maximum of the signal to noise ratios and its 
approximation for the case of accumulation periods lengths of 1~s and 0.1~s 
are presented in figures~\ref{f:prob_snr_long} and \ref{f:prob_snr_short}. 

  Using effective parameters of the a~priori distribution, we can compute the
probability that a given observation with the signal to noise ratio $s$ 
belongs to the population of observations without signal from a source as
\beq
    N(s) = 1 - \ds\int\limits_0^s p(s) \, ds = 
           1 - \frac{1}{<\!\!z_{\eff}\!\!>}\left(1 - 
                         e^{-\frac{s^2}{\pi}} \right)^{n_{\tiny\eff}}
\eeq{e:e4}

  Examining plots of the a~posteriori distribution density, we conclude that
at the range of signal to noise ratios [3.5, 6.5] the share of points which
belongs to detected sources is insignificant. Neglecting the population of
observations with detected sources slightly inflated our estimates of the
probability that an observation belongs to the population of data without
signal and makes our estimates of the probability of detection a bit more 
conservative.

  Each source was observed $m$ times, $m$ typically being in the range 3--6.
At a given series of signal to noise ratios, a source has $k$ observations 
with signal to noise ratio greater than $s_{lim}$. The probability of 
false detection $p_{lim}$ corresponds to it. Then for the case when
$p_{lim}$ is small that we can neglect terms of $p_{lim}^2$, the probability 
that this may happen due only to the noise in the data, i.e. a source was not
detected in neither of $k$ observations, is
\beq
      N_k = C^k_m \ds\prod_{i=1}^{i=k} N(s_i)
\eeq{e:e5}

   We used this expression in order to compute the probability of false
detection for each source. The value $p_{lim} = 0.05 $ was used for 
computations.

\section{Results and discussion}
\label{s:results}

   We split the list of 2494 sources into three groups: not detected if the
probability of false detection is higher than $ 10^{-2}$; marginally detected
if the probability of false detection is in the range of 
$[10^{-5},\, 10^{-2}]$; and reliably detected if this probability is less
than $10^{-5}$. 

  Careful analysis revealed that 16 detected sources are known water masers. 
The H$_2$O line was within the recorded bandwidth. We made a trial 
fringe search using only one frequency channel which included the water maser 
line. The fringe amplitude {\it decreased} by factor of 4 for continuum 
sources and {\it increased} by the same factor for maser sources.

  The catalogue of 533 detected sources, which are not masers, is listed in 
table~\ref{t:source_detected}.
Source coordinates are taken from the NVSS catalogue \citep{nvss}. 
Column /5/ shows the detection status: {\tt D} --- reliably detected, 
{\tt M} --- marginally detected. The correlated flux density in jansky is at 
three ranges of baseline projections are given in columns /6/, /7/, /8/: in the 
range of [0, 70] megawavelengths; in the range of [70, 100] megawavelengths; 
and in the range of [100, 200] megawavelengths. Column /9/ gives the 
probability of false detection, if it exceeds $10^{-5}$, or zero, if it is 
less than $10^{-5}$. Column /10/ gives a flag denoting whether the source
was previously detected in the X/S surveys ({\tt KNO}) or not ({\tt NEW}).

\ifpreprint
\begin{deluxetable*}{r r r r r r r r r r }
\else
\begin{deluxetable}{r r r r r r r r r r }
\fi
   \tablecaption{List of detected continuum spectrum sources 
                 \label{t:source_detected}}
   \tablehead{
          J2000 name         &
          B1950 name         &
          $ \alpha $ (J2000) &
          $ \delta $ (J2000) &
          \ntab{c}{(5)}      & 
          \ntab{c}{(6)}      & 
          \ntab{c}{(7)}      & 
          \ntab{c}{(8)}      & 
          \ntab{c}{(9)}      & 
          \ntab{c}{(10)} 
    }
    \startdata
\tt J0001+6051 & \tt 2358+605 & \tt 00:01:07.09 & \tt +60:51:22.8 & D & \nodata & 0.16    & 0.21    & 0.0    & \tt KNO \\
\tt J0005+5428 & \tt 0002+541 & \tt 00:05:04.36 & \tt +54:28:24.9 & D & \nodata & 0.32    & 0.28    & 0.0    & \tt KNO \\
\tt J0006+5050 & \tt 0003+505 & \tt 00:06:08.29 & \tt +50:50:03.4 & D & \nodata & \nodata & 0.17    & 0.0    & \tt NEW \\
\tt J0014+6117 & \tt 0012+610 & \tt 00:14:48.79 & \tt +61:17:43.5 & D & 0.29    & 0.25    & 0.33    & 0.0    & \tt KNO \\
\tt J0019+2021 & \tt 0017+200 & \tt 00:19:37.85 & \tt +20:21:45.6 & D & \nodata & 0.89    & 0.91    & 0.0    & \tt KNO \\
\tt J0019+7327 & \tt 0016+731 & \tt 00:19:45.78 & \tt +73:27:30.0 & D & 1.24    & 1.35    & 1.28    & 0.0    & \tt KNO \\
\tt J0021-1910 & \tt 0018-194 & \tt 00:21:09.37 & \tt -19:10:21.3 & M & 0.53    & 0.09    & \nodata & 5.D-03 & \tt NEW \\
\tt J0027+5958 & \tt 0024+597 & \tt 00:27:03.28 & \tt +59:58:52.9 & D & 0.35    & 0.49    & 0.28    & 0.0    & \tt KNO \\
\tt J0037-2145 & \tt 0034-220 & \tt 00:37:14.79 & \tt -21:45:24.5 & D & \nodata & 0.33    & 0.12    & 5.D-06 & \tt NEW \\
    \enddata
\tablecomments{Table~\ref{t:source_detected} is presented in its entirety 
               in the electronic edition of the Astronomical Journal. 
               A portion is shown here for guidance regarding its form and 
               contents. Units of right ascension are hours, minutes and 
               seconds; units of declination are degrees, minutes and seconds;
               units of flux density is jansky.}
\ifpreprint
\end{deluxetable*}
\else
\end{deluxetable}
\fi

  Table \ref{t:source_not_detected} lists 1945 non-detected sources. Columns
/5/, /6/ and /7/ provide the minimum, average and maximal upper limit
of the correlated flux density. These quantities are computed on the basis
of the calibrated gain of the interferometer considering what correlated
flux density would provide a signal to noise ratio of 6.0. Column /8/ gives
the number of observations used in the analysis, and column /9/ gives 
a flag denoting whether the source was previously detected at the X/S survey 
({\tt KNO}) or not ({\tt NEW}).

\ifpreprint
\begin{deluxetable*}{r r r r r r r r r }
\else
\begin{deluxetable*}{r r r r r r r r r}
\fi
   \tablecaption{List of not detected sources \label{t:source_not_detected}}
   \tablehead{
          J2000 name         &
          B1950 name         &
          $ \alpha $ (J2000) &
          $ \delta $ (J2000) &
          \ntab{c}{(5)}      & 
          \ntab{c}{(6)}      & 
          \ntab{c}{(7)}      & 
          \ntab{c}{(8)}      & 
          \ntab{c}{(9)}       
    }
    \startdata
\tt J0000+3252 & \tt 2358+326 & \tt 00:00:49.72 & \tt +32:52:56.5 &  0.11 &  0.15 &  0.18 &   3 & \tt NEW \\
\tt J0000+5539 & \tt 2357+553 & \tt 00:00:20.45 & \tt +55:39:08.6 &  0.13 &  0.15 &  0.16 &   3 & \tt NEW \\
\tt J0001+6443 & \tt 2358+644 & \tt 00:01:14.95 & \tt +64:43:01.1 &  0.20 &  0.33 &  0.88 &   9 & \tt NEW \\
\tt J0002+5510 & \tt 2359+548 & \tt 00:02:00.47 & \tt +55:10:38.0 &  0.15 &  0.15 &  0.16 &   2 & \tt NEW \\ 
%
    \enddata
\tablecomments{Table~\ref{t:source_not_detected} is presented in its entirety 
               in the electronic edition of the Astronomical Journal. 
               A portion is shown here for guidance regarding its form and 
               contents. Units of right ascension are hours, minutes and 
               seconds; units of declination are degrees, minutes and seconds;
               unit of flux density is jansky.}
\ifpreprint
\end{deluxetable*}
\else
\end{deluxetable}
\fi

\begin{deluxetable}{l r r}
   \tablecaption{Statistics of observed sources \label{t:source_table}}
   \tablehead{
          Category &
          Detected &
          Not detected \\
    }
    \startdata
      New, continuum spectrum             & 180  & 1721    \\
      Known, water masers                 &  16  & \nodata \\
      Known X/S,   continuum spectrum     & 353  &  224    \\
      Known X/S/K, amplitude calibrators  &  73  &    3    \\
      \hline
      Total                               & 549  & 1945    \\
   \enddata
\end{deluxetable}

  The statistics of the sample of observed sources is given in table 
\ref{t:source_table}. VERA detected $~60$\% of the known X/S sources and 
$~10$\% of the new sources. These results are somewhat disappointing. With 
the best weather conditions VERA is able to detect sources as weak as 100~mJy
with 2~minute integration times. However, even in winter, adverse weather 
conditions significantly deteriorated the detection limit. We show in 
figure~\ref{f:prob_ndt} the cumulative probability function that a source 
will have a fringe amplitude signal to ratio limit greater than 6.0 among 
14 observing sessions. This detection limit was computed on the basis of the 
calibrated gain. This probability distribution characterizes the current 
averaged VERA single beam sensitivity in winter when observations are 
carried out with the integration times of 120~s. 

\begin{figure}[ht]
  \par\vspace{4ex}
  {\includegraphics[width=0.48\textwidth,clip]{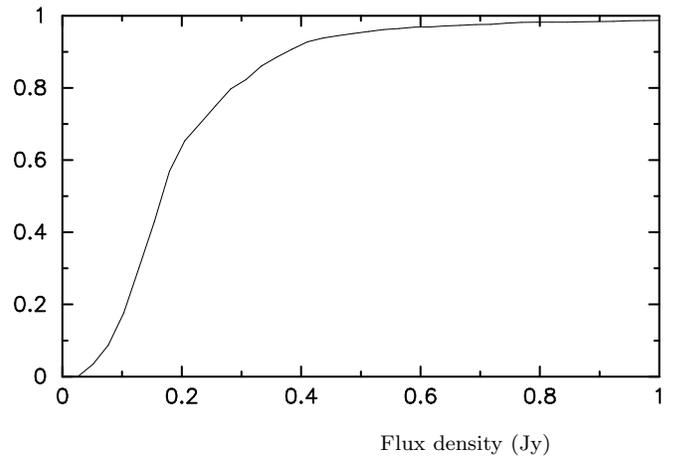} }
  \par\vspace{-0.1ex}\par\noindent\hspace{49mm}{Flux density (Jy)}
  \caption{Cumulative probability function for a source to have signal 
           to noise ratio above the detection limit 6.0 over 14 observing 
           sessions with VERA at 22~GHz.}
  \par\vspace{-4ex}\par\hphantom{a}
  \label{f:prob_ndt}
  \par\vspace{-4ex}
\end{figure}

   Since the 2005f\_astro X/S catalogue contains 3297 objects within the
declination zone $ > -40^\circ$, the zone which the VERA can observe, and 
approximately 60\% of the sources from this catalogue were detected, we can 
roughly estimate that the VERA can detect $~2000$ sources from that list. 
To date, 252 sources were detected at K-band with the VLBA in the K/Q survey, 
280 other known X/S calibrators and 180 new sources were detected with 
the VERA in this campaign, and 64 sources were detected with VERA at various
other experiments. In total, the pool of confirmed K~band calibrators 
is 776 objects.


\section{Concluding remarks}
\label{s:summary}

   In the 2005/2006 winter campaign $\sim 2500$ sources were observed with 
the VERA during 148 hours. The target sources were either within $6^\circ$ 
of the Galactic plane, or within $11^\circ$ of the Galactic center,
or within $2.\!{}^\circ 2$ of SiO or H$_2$O masers. Among them, 
533 continuum spectrum and 16 maser sources were detected, including 180 
continuum spectrum sources not previously observed with VLBI. The estimates 
of the correlated flux densities in the ranges [0, 70], [70, 100], 
and [100, 200] megawavelengths were evaluated. For 1945 non-detected sources 
the upper limits of their correlated flux densities were evaluated. 

  The distribution of detected sources over the sky is shown in 
figure~\ref{f:sou_map}. All detected sources are scheduled for observations
with the VLBA at 22~GHz in June-October 2006 for imaging and determination of
their positions with 1--10~nanoradian accuracies. Tables
of detected and non-detected sources, plots of system temperatures, 
fringe plots, and other auxiliary information related to this campaign
can be found on the Web at {\tt http://lacerta.gsfc.nasa.gov/vlbi/fss}.

\begin{figure*}[ht]
  {\includegraphics[width=\textwidth,clip]{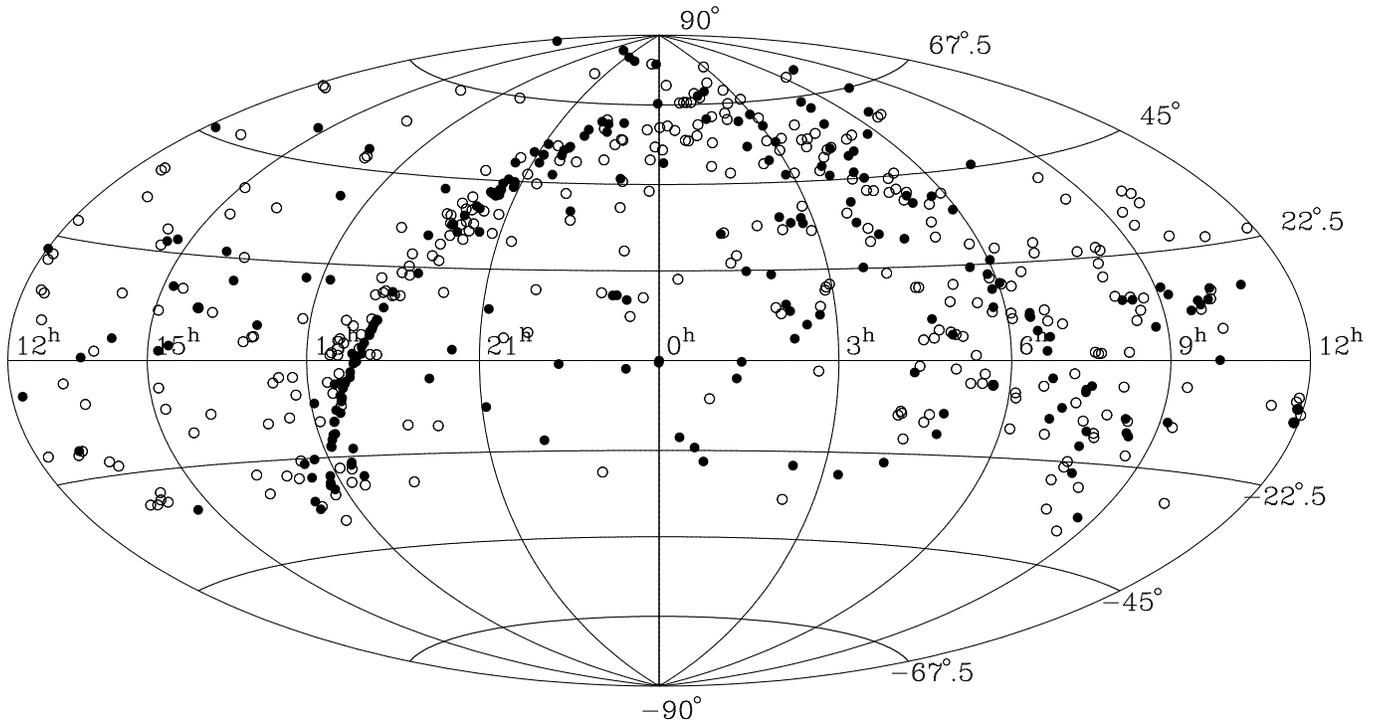} }
  \caption{The distribution of detected sources over the sky. Disks designate
           new compact source not observed before with VLBI, circles 
           designate known sources observed before at X-band with the VLBA.}
  \label{f:sou_map}
\end{figure*}

  Analysis of the VERA fringe survey data shows that the probability of 
detecting a 100~mJy source with 120~s of integration time is about 10\%;
the probability of detecting a 200~mJy source is 60\%; and the probability 
of detecting a 300~mJy source is 80\%. Future development a next generation
water vapor radiometer (Kawaguchi (2006), private communication) promises 
to allow increasing the integration time and, thus, to improve VERA 
sensitivity.

\acknowledgments

The authors are thankful to D.~Gordon for valuable comments. The authors made 
use of the database CATS \citep{cats} of the Special Astrophysical 
Observatory. This research has made use of the NASA/IPAC Extragalactic 
Database (NED) which is operated by the Jet Propulsion Laboratory, California 
Institute of Technology, under contract with the National Aeronautics and 
Space Administration.

\end{document}